\documentclass[11pt,letterpaper]{article}

\usepackage[letterpaper,margin=1in]{geometry}
\usepackage{amsmath,amssymb,amsthm}
\usepackage{tikz}
\usepackage{thmtools}
\usepackage{thm-restate}
\usetikzlibrary{decorations.pathreplacing,fadings}
\usepackage[hidelinks]{hyperref}

\declaretheorem[name=Theorem]{theorem}
\declaretheorem[name=Lemma,sibling=theorem]{lemma}
\declaretheorem[name=Corollary,sibling=theorem]{corollary}
\newcommand{\OPT}{\operatorname{OPT}}
\newcommand{\Anc}{\operatorname{Anc}}
\newcommand{\ADSO}{\operatorname{ADSO}}

\newcounter{algorithm}
\newenvironment{algorithm}[1]{%
  \refstepcounter{algorithm}\par\medskip\noindent
  \begin{minipage}{\linewidth}\hrule\smallskip
  \textbf{Algorithm \thealgorithm: #1}\par\smallskip\hrule\medskip
  \normalsize
}{%
  \smallskip\hrule\end{minipage}\par\medskip
}

\title{Approximating Submodular Orienteering\\%
       via Rooted Subgraph Decomposition}
\author{%
  Lars Rohwedder\thanks{%
    University of Southern Denmark (SDU). Email:
    \href{mailto:rohwedder@sdu.dk}{\texttt{rohwedder@sdu.dk}}.%
  }%
}
\date{}

\begin{document}
\hypersetup{pageanchor=false}
\begin{titlepage}
  \maketitle
  \thispagestyle{empty}

  \begin{abstract}
	  \normalsize
	  The Submodular Orienteering problem asks for an $s$--$t$ walk in a directed graph
	  that maximizes a monotone submodular function over the visited vertices subject to a bound on its length.
    We give a deterministic polynomial-time approximation algorithm with subpolynomial approximation rate
    $2^{O(\sqrt{\log n\,\log\log n})}$.
    This improves over the $O(n^\varepsilon)$ approximation rate
    that was previously known for every fixed $\varepsilon>0$.

    We prove a decomposition lemma that partitions a directed graph into rooted
    subgraphs such that every path intersects only polylogarithmically many
    such subgraphs. Our algorithm uses this decomposition to construct its recursive
    subproblems.
  \end{abstract}
\end{titlepage}
\setcounter{page}{1}
\hypersetup{pageanchor=true}

\section{Introduction}

In Submodular Orienteering we are given a directed graph $G=(V,A)$ with
$n=|V|$, vertices $s,t\in V$, a length function $\ell:A\to\mathbb Q_{\ge0}$, a
budget $C\in\mathbb Q_{\ge0}$, and a nonnegative monotone submodular function
$f$ on the vertices. The goal is to find an $s$--$t$ walk $P$ of total length
$\ell(P)\le C$ maximizing $f(P)$, the value of its visited vertex set. Write
$\OPT$ for the maximum reward of a feasible walk. Access to $f$ is via a value
oracle whose answers have polynomial encoding length.

Informally, Submodular Orienteering models routing with a limited travel
budget and diminishing returns from visiting additional locations. Since its
introduction by Chekuri and P\'al~\cite{CP05}, the problem has motivated
algorithms for example in environmental monitoring~\cite{SKGK09},
robotic search-and-rescue applications~\cite{SKK09}, and personalized route recommendation~\cite{ZCC15}. In
reinforcement learning with diminishing returns, submodular rewards are used to capture
the value of an entire trajectory of a Markov decision
process~\cite{WZCGS20,PMZK24,DPK24}. This is again closely connected to
Submodular Orienteering, see~\cite{RZ26}.

The best previously known polynomial-time approximation for arbitrary monotone
submodular rewards is $O(n^\varepsilon)$ for every fixed $\varepsilon>0$, due
to Rohwedder and Zenklusen~\cite{RZ26}; their running time is
$n^{O(1/\varepsilon)}$. In quasi-polynomial time, the Recursive Greedy
algorithm of Chekuri and P\'al~\cite{CP05} gives an $O(\log n)$ approximation.
They also show a nearly matching $\Omega(\log^{1-\varepsilon}n)$ hardness
bound for every fixed $\varepsilon>0$, unless $\mathrm{NP}$ admits Las Vegas
algorithms with quasi-polynomial expected running time. The special case in
which $f$ counts the distinct colors of visited vertices has been
independently studied by Cou\"etoux, Nakache, and Vax\`es~\cite{CNV17}, who
obtain a different $O(n^\varepsilon)$ approximation and prove that
no constant-factor approximation exists unless $\mathrm P=\mathrm{NP}$.

A very similar gap between polynomial and quasi-polynomial time occurs in
Directed Steiner Tree~\cite{CCC99,GLL19}, general additive and submodular
Santa Claus~\cite{CCK09,BMR25}, Robust $s$--$t$ Path~\cite{LXZ24},
Robust Bipartite Perfect Matching, as well as a number of other problems,
see~\cite{BLR26}. These problems have $O(n^\varepsilon)$ approximations in
time $n^{O(1/\varepsilon)}$ for every fixed $\varepsilon>0$, and
polylogarithmic approximations in quasi-polynomial time, where $n$ denotes the
respective input size. The quasi-polynomial regime can be viewed as choosing a
subconstant $\varepsilon$, for example
$\varepsilon=\Theta(\log\log n/\log n)$: the approximation rate becomes
polylogarithmic, while the running-time exponent becomes superconstant.
The algorithmic techniques in these different problems are very similar, typically based
on either an idea similar to Recursive Greedy or randomized rounding on an
LP relaxation with some recursive structure.
These common features make it particularly meaningful to obtain a subpolynomial
approximation in polynomial time for any of these problems.

We obtain such a guarantee for Submodular Orienteering.

\begin{theorem}
  \label{thm:main}
  There is a deterministic polynomial-time algorithm that, for every feasible
  instance of Submodular Orienteering, returns an $s$--$t$ walk $P$ respecting
  the length budget such that
  \[
    f(P)\ge2^{-O(\sqrt{\log n\,\log\log n})}\OPT.
  \]
\end{theorem}


A reduction in~\cite[Theorem~3]{CRSVZ26} maps Submodular Matroid
Intersection Basis to Submodular Orienteering with only a constant-factor
loss. Thus, Theorem~\ref{thm:main} implies the following.

\begin{samepage}
\begin{corollary}
  \label{cor:matroid}
  Let $\mathcal M_1=(E,\mathcal I_1)$ and $\mathcal M_2=(E,\mathcal I_2)$ be
  matroids given by independence oracles, with $n=|E|$. For every feasible
  cardinality $K$ and nonnegative monotone submodular function $f$ given by a
  value oracle, a deterministic polynomial-time algorithm returns
  $S\in\mathcal I_1\cap\mathcal I_2$ with $|S|=K$ and
  \[
    f(S)\ge2^{-O(\sqrt{\log n\,\log\log n})}
    \max_{\substack{T\in\mathcal I_1\cap\mathcal I_2\\|T|=K}}f(T).
  \]
  In particular, the same approximation rate holds for matroid intersection bases and 
  bipartite perfect matchings.
\end{corollary}
\end{samepage}

\subsection{Technical overview and comparison to previous algorithms}
\label{sec:comparison}

We show that by preprocessing, one can simplify the graph to a DAG
and omit the length constraint; hence, feasible solutions are exactly the $s$--$t$ paths.
One may further assume that all $s$--$t$ paths have the same number of arcs, which simplifies
the Recursive Greedy technique below, but it is not necessary for our final algorithm.
We start by summarizing two important previous techniques. 

Chekuri and P\'al~\cite{CP05} introduced the elegant Recursive Greedy algorithm.
It guesses the middle vertex of the path, recursively constructs
a prefix ending at that vertex, and then optimizes the suffix with respect to
the prefix's marginal reward. This gives a logarithmic approximation in
quasi-polynomial time. The running time comes from the fact that the algorithm makes polynomially many
recursive calls per invocation and
the depth of the recursion is $O(\log n)$, since the number of arcs on $s$--$t$ paths is halved
by each recursion.

Rohwedder and Zenklusen~\cite[Appendix~A]{RZ26}, based on unpublished work of
Fischer and Rohwedder, generalize Recursive Greedy and
obtain more flexible tradeoffs between running time and approximation. 
In their algorithm the recursion depth can be set arbitrarily. At depth $O(1/\varepsilon)$,
they obtain approximation rate $O(1/\varepsilon \cdot n^{\varepsilon})$ and running time $n^{O(1/\varepsilon)}$.
This is achieved by splitting the optimal path (more precisely, the graph) not just into two parts
for prefix and suffix, but into possibly more parts.
To avoid a blowup in the running time, the algorithm guesses only endpoints of one segment at a time,
which is achieved using dynamic programming techniques.
Rohwedder and Zenklusen also give an LP-based algorithm that extends
to submodular Markov decision processes and achieves similar running time and approximation rate tradeoffs.

For this overview, we will gradually develop the approach used in this paper. We start with simpler, but
flawed approaches, and thereby motivate the more involved ideas.

The bottleneck of Recursive Greedy and the variant by Rohwedder and Zenklusen is the
guessing step and therefore our goal is to avoid it. First, consider Recursive Greedy in an all-destinations variant, where
we compute a good path from the source to every other vertex of the graph. This allows us to combine the recursive
calls for the prefix into one; constructing the suffix still requires one call for each guess.
The problem there is that each guessed middle point (the source in the recursion) leads to a different marginal reward
function and therefore to incompatible subproblems.
We prove the following structural lemma that will help us avoid this:
\begin{samepage}
\begin{restatable}{lemma}{partition}
  \label{lem:partition}
  Let $G=(V,A)$ be a directed graph with $n=|V|$. In deterministic
  polynomial time we can construct a partition $\mathcal P$ of $V$
  such that:
  \begin{enumerate}
    \item Each part $U\in\mathcal P$ contains a root $r_U$
      from which every vertex in $U$ can be reached within $G[U]$.
    \item Every directed path intersects $O(\log^2 n)$ parts.
  \end{enumerate}
\end{restatable}
\end{samepage}
A first idea would be to apply it directly for the suffixes. (The idea is flawed and should not be taken as a precise argument.) Assuming all $s$--$t$ paths have the same number of arcs,
we restrict to the region that is relevant for the suffixes, namely the latter half of the graph,
and decompose the resulting graph using the lemma above. The roots of the subgraphs should replace the 
guessed middle points in the suffix recursions. We optimize in each subgraph separately over the paths from its root to all destinations.
For the objective function, we take the marginal reward with respect to the best path from the prefix recursion
that can reach the subgraph's root.
We then output the best solution over all subgraphs, where the solution includes the corresponding prefix, the path within the subgraph
and the paths connecting both paths and connecting the latter path to $t$. In terms of running time, this algorithm is much more economical, because each vertex
appears in only one of the suffix recursions (and in the prefix recursion), whereas in Recursive Greedy
vertices can appear in many recursions. One can hope to argue that choosing the best of the solutions leads to a loss
of at most a factor of $O(\log^2 n)$ in the approximation rate, using that this many subgraphs intersect
the optimal path.
Crucially, this loss may happen for each level of the recursion tree.
This alone might still be manageable, because with the Rohwedder-Zenklusen variant we could potentially
reduce the recursion depth to, say, $\sqrt{\log n}$, which keeps this loss at $(\log n)^{O(\sqrt{\log n})}$, i.e., subpolynomial.

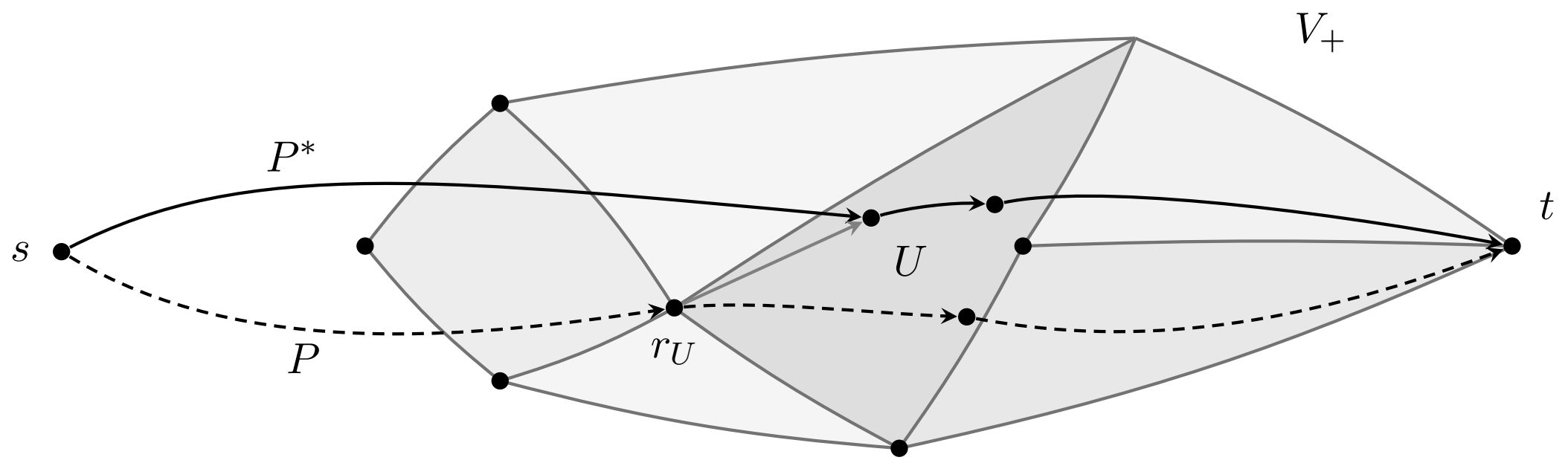
\begin{figure}
  \centering
\begin{tikzpicture}[x=1cm,y=1cm,>=stealth,
    every node/.style={font=\normalsize},
    vertex/.style={circle,fill=black,inner sep=1.55pt},
    boundary/.style={thick,black!55}]

  \coordinate (A) at (2.70,0.05);
  \coordinate (B) at (3.90,1.32);
  \coordinate (C) at (9.55,1.90);
  \coordinate (T) at (12.90,0.05);
  \coordinate (E) at (7.45,-1.75);
  \coordinate (F) at (3.90,-1.15);
  \coordinate (R) at (5.45,-0.50);
  \coordinate (X) at (8.55,0.05);

  \fill[black!7] (A) to[bend left=6] (B)
    to[bend left=8] (R) to[bend left=6] (F)
    to[bend left=6] (A) -- cycle;
  \fill[black!4] (B) to[bend left=4] (C)
    to[bend right=3] (R) to[bend right=8] (B) -- cycle;
  \fill[black!13] (R) to[bend left=3] (C)
    to[bend left=5] (X) to[bend left=4] (E)
    to[bend left=4] (R) -- cycle;
  \fill[black!4] (F) to[bend right=6] (R)
    to[bend right=4] (E) to[bend left=5] (F) -- cycle;
  \fill[black!5] (X) to[bend right=5] (C)
    to[bend left=6] (T) to[bend right=2] (X) -- cycle;
  \fill[black!9] (E) to[bend right=4] (X)
    to[bend left=2] (T) to[bend left=6] (E) -- cycle;

  \foreach \from/\to/\bend in {
      A/B/6,B/C/4,C/T/6,T/E/6,E/F/5,F/A/6,
      B/R/8,R/F/6,R/C/3,C/X/5,X/E/4,E/R/4,X/T/2}
    \draw[boundary] (\from) to[bend left=\bend] (\to);
  \node at (11.20,1.95) {$V_+$};
  \node at (7.55,-0.08) {$U$};

  \foreach \name in {A,B,E,F,X}
    \node[vertex] at (\name) {};
  \node[vertex] (s) at (0,0) {};
  \node[vertex] (r) at (R) {};
  \node[vertex] (u) at (7.20,0.30) {};
  \node[vertex] (v) at (8.30,0.42) {};
  \node[vertex] (w) at (8.05,-0.58) {};
  \node[vertex] (t) at (T) {};

  \draw[->,thick] (s)
    .. controls (1.65,0.85) and (3.40,0.65) .. (u);
  \draw[->,thick] (u)
    .. controls (7.55,0.40) and (7.95,0.44) .. (v);
  \draw[->,thick] (v)
    .. controls (9.15,0.60) and (11.00,0.40) .. (t);
  \node at (2.05,0.85) {$P^*$};

  \draw[->,thick,densely dashed] (s)
    .. controls (1.65,-1.00) and (3.75,-0.75) .. (r);
  \draw[->,thick,densely dashed] (r)
    .. controls (6.25,-0.43) and (7.20,-0.55) .. (w);
  \draw[->,thick,densely dashed] (w)
    .. controls (9.50,-0.85) and (10.90,-0.70) .. (t);
  \node at (2.15,-0.95) {$P$};

  \draw[->,thick,black!50] (r)--(u);
  \node[left=4pt] at (s) {$s$};
  \node[below=4pt] at (r) {$r_U$};
  \node[above right=3pt] at (t) {$t$};
\end{tikzpicture}
\caption{Application of the rooted subgraph decomposition lemma with optimal path $P^*$ and algorithm's output $P$}
\label{fig:decomp}
\end{figure}
Another, more crucial issue is that
the optimal prefix may not be able to reach the root of the subgraph with the highest value (the optimal
path may enter it through another vertex). Hence, one might lose all of the optimal prefix's value by
having to take an entirely different path that can reach the subgraph's root.
This makes additional modifications necessary.
Recall that Recursive Greedy and the Rohwedder-Zenklusen variant make progress by splitting the optimal path
into smaller segments.
We will split the path not by number of arcs, but by objective value.
We consider the prefix of the optimal path that constitutes a $(1 - \lambda)$-fraction
of the objective value and the suffix that constitutes the remaining $\lambda$-fraction in marginal value.
First, we recursively try to find paths of value $(1 - \lambda)\OPT / \alpha$ to all vertices,
where $\alpha$ is an approximation rate. Let $V_+$ be those where we succeed.
Then we apply the decomposition lemma on $G[V_+]$ and follow an approach similar to above, see also Figure~\ref{fig:decomp}.
The roots now have by definition a path of high value leading to them, resolving the previous issue.
Through the recursions, the target objective value decreases and
by a simple preprocessing step one can establish that there are no tiny non-zero objective values.
This ensures that
the recursion terminates quickly for the right choice of $\lambda$.

For future reference, we summarize the precise properties that we establish through preprocessing 
in the following lemma.
\begin{samepage}
\begin{restatable}{lemma}{preprocessing}
  \label{lem:preprocessing}
  We can in polynomial time transform a feasible instance $I$ of Submodular Orienteering
	into instances $I_1,\dotsc,I_k$ that each satisfy:
	\begin{enumerate}
		\item The number of vertices in $I_i$ is polynomially bounded in the number of vertices in $I$.
		\item There is no length bound (equivalently, $C = \infty$).
		\item The input graph is a transitively closed DAG, where each vertex lies on some $s$--$t$ path.
		\item The marginal rewards satisfy
\begin{equation*}
	f(\{v\}\mid S)\ge\frac{f(V)}{n^2}
	\qquad\forall S\subseteq V,\ v\in V\setminus S. \footnote{Here, $V, n$, and $f$ refer to instance $I_i$.}
\end{equation*}
	\end{enumerate}
	Given $\alpha$-approximations for $I_1,\dotsc,I_k$, we can in polynomial time compute
	a $6\alpha$-approximation for~$I$.
\end{restatable}
\end{samepage}

\subsection{Other related work}

With a hard length constraint, undirected Submodular Orienteering is
approximation-equivalent to its directed version. This follows from a
reduction to an $(H+1)$-layered DAG.
A reduction similar to that in Section~\ref{sec:length} can construct such an instance. 
For an instance with budget~$B$,
forget the arc directions, add $B+1$ to every length, and set the budget to
$H(B+1)+B$. Any feasible undirected $s$--$t$ walk must go closer to $t$ in every step,
since a backward step would exceed the budget. Thus it is a
feasible directed path with the same reward. The reverse direction simply
replaces each undirected edge with two oppositely directed arcs.

The additive special case of Orienteering admits stronger approximation
guarantees. Chekuri, Korula, and P\'al~\cite{CKP12} give a polynomial-time
$(2+\varepsilon)$-approximation in undirected graphs and
a polynomial-time $O(\log^2 n)$-approximation in directed graphs. 
Nagarajan and Ravi~\cite{NR11} achieved further improvements for directed graphs,
but obtaining a constant-factor approximation remains an interesting open question, see~\cite[Open Problem~17.9]{TV24}.

Our approach applies to additive rewards but does not improve these
guarantees. The reduction in Section~\ref{sec:length} replaces each vertex by
several copies, whose reward must be counted only once; the resulting coverage
objective is submodular rather than additive and therefore loses most of the additional structure.

C\u{a}linescu and Zelikovsky~\cite{CZ05} study Polymatroid Steiner problems,
including directed Polymatroid Steiner Tree. Given a directed graph with
nonnegative arc costs, a root, and a normalized integer-valued monotone
submodular rank function $f$, the goal is to find a minimum-cost rooted
arborescence $T$ with $f(V(T))=f(V)$. This generalizes Directed
Steiner Tree, Group Steiner Tree, and Covering Steiner Tree. When $f(V)$ is polynomially bounded in $n$, 
they obtain polylogarithmic approximations in quasi-polynomial time
and an $O(n^\epsilon)$-approximation in polynomial time.

Ghuge and Nagarajan~\cite{GN20} study the related Submodular Tree Orienteering, which asks
for a rooted arborescence maximizing a monotone submodular reward subject to a
cost budget. They extend Recursive
Greedy to obtain an $O(\log n/\log\log n)$ approximation in quasi-polynomial
time.

\subsection{Notation}
\label{sec:preliminaries}

\paragraph{Submodular functions.}
A function $f:2^V\to\mathbb R_{\ge0}$ is monotone if $f(X)\le f(Y)$ for
$X\subseteq Y$, and submodular if $f(X)+f(Y)\ge f(X\cup Y)+f(X\cap Y)$ for all
$X,Y\subseteq V$. We normalize $f(\emptyset)=0$; a nonzero value at the empty
set can be subtracted and added back afterward. In particular, $f$ is
subadditive: $f(X\cup Y)\le f(X)+f(Y)$. For sets or paths, write
\[
  f_Y(X)=f(X\mid Y)=f(X\cup Y)-f(Y),
\]
interpreting a path as its vertex set. Marginal functions are again
normalized, monotone, and submodular. 

\paragraph{Graphs and paths.}
For $U\subseteq V$, write $G[U]$ for the subgraph induced by $U$. Reachability
includes the length-zero path from a vertex to itself. A directed graph is
transitively closed if, whenever $v$ reaches another vertex $u$, the graph
contains the arc $(v,u)$. Paths are nonempty and may consist of one vertex.
For a path $P$, write $V(P)$ and $A(P)$ for its vertex and arc sets,
respectively. For a walk $P$, its length $\ell(P)$ sums arc lengths with
multiplicity. If the graph is transitively closed, a nonempty intersection
$P\cap U$ with $U\subseteq V$ is viewed as the path on those vertices in their
original order. We write $P\circ P'$ for the concatenation of $P$ and $P'$,
adding a connecting arc if necessary, whenever the result is a path.
\bigskip

\noindent
Unless stated otherwise, $\log$ has base $2$ and $\ln$ denotes the natural logarithm.

\subsection{Organisation}
Section~\ref{sec:preliminaries} gives definitions and notation.
Section~\ref{sec:algorithm} 
proves a simpler randomized version of the decomposition lemma and gives the algorithm for
DAGs without length constraint.
Sections~\ref{sec:guarantee} and~\ref{sec:runtime} prove the approximation
guarantee and running time. Section~\ref{sec:length} proves the instance simplification
as in Lemma~\ref{lem:preprocessing}. Section~\ref{sec:decomposition} gives
the full deterministic construction of the decomposition lemma.

\section{Algorithm}
\label{sec:algorithm}

We describe the algorithm for a simplified instance with the assumptions as in Lemma~\ref{lem:preprocessing}.
Recall that since the input is a DAG, each walk is a path.
We assume that $n>1$ and $f(V)>0$. Otherwise, the problem is trivial. 

For intuition, we give a simple proof for a randomized version of the decomposition lemma with the stronger bound
$O(\log n)$ in expectation for each fixed path. Such a randomized construction could
already be used in the algorithm to obtain a randomized approximation
guarantee; we omit the details. The proof of the full deterministic lemma is
deferred to Section~\ref{sec:decomposition}.

\begin{proof}[Proof of randomized version of Lemma~\ref{lem:partition}]
  We use Cohen's least-element mapping~\cite{Coh97}; each fixed path will
  intersect $O(\log n)$ parts in expectation.

  For each vertex $v$, let $\Anc(v)$ consist of $v$ and all vertices that can
  reach it. Choose a uniformly random ordering of $V$, let each vertex choose
  its first ancestor in this ordering, and group vertices making the same
  choice. If $v$ chooses $u$, every vertex $w$ on a $u$--$v$ path also chooses
  $u$: the vertex $u$ reaches $w$, and any earlier ancestor of $w$ would also
  reach $v$. Thus each part contains its chosen root and paths from that root
  to all its vertices, as required.

  Fix a path $P=(u_1,\ldots,u_h)$ before sampling. The vertices of $P$
  reachable from any ancestor of $u_h$ form a suffix. List the ancestors of
  $u_h$ as $v_1,\ldots,v_k$ in nonincreasing order of the number of vertices
  of $P$ they can reach, breaking ties in a fixed way. If a path vertex
  chooses $v_i$, all of $v_1,\ldots,v_i$ reach it, so $v_i$ must be first
  among them in the random ordering. This necessary event has probability
  $1/i$, and hence
  \[
    \mathbb E\bigl[|\{U\in\mathcal P:U\cap V(P)\ne\emptyset\}|\bigr]
    \le\sum_{i=1}^k\frac1i\le1+\ln n. \qedhere
  \]
\end{proof}
The deterministic construction in Section~\ref{sec:decomposition} gives a
simultaneous bound of $O(\log^2 n)$ parts for all paths.
We now sketch how the decomposition can be used in a recursion towards the
claimed approximation guarantee, giving more technical details than in the introduction.
We defer running time considerations to the
end of the sketch. We slightly oversimplify the
recursive guarantees and describe progress as a decrease in the optimum; the
full analysis makes these statements precise.

Fix an optimal $s$--$t$ path $P^*=(v_1,\ldots,v_h)$ of value $\OPT>0$. Choose
$\lambda=2^{-\Theta(\sqrt{\log n\,\log\log n})}$, and let $i\in\{1,\ldots,h\}$
be minimal such that
\[
  f(\{v_1,\ldots,v_i\})>(1-\lambda)\OPT.
\]
For a desired approximation rate $\alpha\ge1/\lambda$, our goal is to find a
path of value at least $\OPT/\alpha$. Imagine that a first recursive call, a \emph{prefix recursion},
has already found paths from $s$ of value at least
$(1-\lambda)\OPT/\alpha$ ending at every vertex admitting a path of value
greater than $(1-\lambda)\OPT$. Let $V_+$ be the vertices with stored path
value at least $(1-\lambda)\OPT/\alpha$. The set $V_+$ contains
$v_i,\ldots,v_h$.

Apply Lemma~\ref{lem:partition} to $G[V_+]$ and the path $P^*\cap V_+$, which
indeed forms a path by transitive closure. The path intersects $O(\log^2 n)$
parts. Each nonempty $P^*\cap U$ is a path in $G[U]$, again by transitive
closure. The possibly empty prefix $P_0^*=(v_1,\ldots,v_{i-1})$ has value at
most $(1-\lambda)\OPT$ by minimality of $i$. The remaining suffix has value at
least $\lambda\OPT$, since by subadditivity the two values sum to at least
$\OPT$. Further, there must be a part $U\in\mathcal P$ with
\begin{equation}
  \label{eq:overview-segment}
  f(P^*\cap U)\ge
  \frac{\lambda\OPT}{O(\log^2 n)}.
\end{equation}
Let $P_U$ be the stored path to the root of this part, of value at least
$(1-\lambda)\OPT/\alpha$. We make a recursive call to optimize the
marginal function $f_{P_U}$ in $G[U]$, which we call a \emph{suffix recursion}. For the sketch, suppose this call
returns a path $R$ from $r_U$ to the last vertex of $P^*\cap U$ with
approximation rate $\alpha'\ge1/\lambda$, so that
\[
  f_{P_U}(R)\ge\frac{f_{P_U}(P^*\cap U)}{\alpha'}.
\]
The concatenation $P_U\circ R$ can be extended to $t$. One can then argue
that for a sufficiently large constant $c$
\begin{equation}
  \label{eq:overview-rate}
  f(P_U\cup R)\ge\frac{\OPT}{\alpha}
  \qquad\text{whenever}\qquad
	\alpha\ge \alpha' \cdot c \log^2 n.
\end{equation}
To get a feeling for why this is true, let us verify the two extreme cases for
$f(P^*\cap U\mid P_U)$.

If $f(P^*\cap U\mid P_U)=0$, then
$f(P_U)\ge f(P^*\cap U)\ge\lambda\OPT/O(\log^2 n)$. Thus, the stored path
already has value at least $\OPT/\alpha$.

\begin{figure}
  \centering
  \begin{tikzpicture}[x=1.1cm,y=0.85cm,>=stealth,
      every node/.style={font=\normalsize}]
    \fill[black!12] (2.85,-0.38) rectangle (5.95,0.38);
    \fill[black!12] (6.65,-1.65) rectangle (8.8,0.38);
    \fill[black!12] (9.55,-0.38) rectangle (11.5,0.38);
    \node at (4.4,0.78) {$U_1$};
    \node at (7.7,0.95) {$U=U_2$};
    \node at (10.5,0.78) {$U_3$};
    \foreach \x/\name in {0/s,1.9/u,3.2/p,4.4/x,5.6/w,7.0/y,8.4/z,9.9/q,11.2/t}
    \node[circle,fill=black,inner sep=1.55pt] (\name) at (\x,0) {};
    \draw[->,thick,densely dotted] (s)--(u);
    \draw[->,thick] (u)--(p);
    \draw[->,thick] (p)--(x);
    \draw[->,thick] (x)--(w);
    \draw[->,thick] (w)--(y);
    \draw[->,thick] (y)--(z);
    \draw[->,thick] (z)--(q);
    \draw[->,thick] (q)--(t);
    \node[above left] at (s) {$s=v_1$};
    \node[below] at (u) {$v_{i-1}$};
    \node[below] at (p) {$v_i$};
    \node[above right] at (t) {$t=v_h$};
    \draw[decorate,decoration={brace,amplitude=4pt}]
      (0,0.7)--(1.9,0.7) node[midway,above=6pt] {$P_0^*$};
    \node[circle,fill=black,inner sep=1.55pt] (v) at (7.7,-1.1) {};
    \draw[->,densely dashed,thick]
      (s) .. controls (0,-1.1) and (1.1,-1.1) .. (2,-1.1)
    -- (v) node[pos=0.42,above] {$P_U$};
    \draw[->,thick,black!50] (v)--(y);
    \draw[->,densely dashed,thick] (v) .. controls (8.35,-1.0) and (8.55,-0.45) .. (z)
      node[pos=0.55,right] {$R$};
    \node[below] at (v) {$r_U$};
    \node[above] at (7.7,0.02) {$P^*\cap U$};
  \end{tikzpicture}
  \caption{Schematic of the recursion}

  \label{fig:optimal-path}
\end{figure}
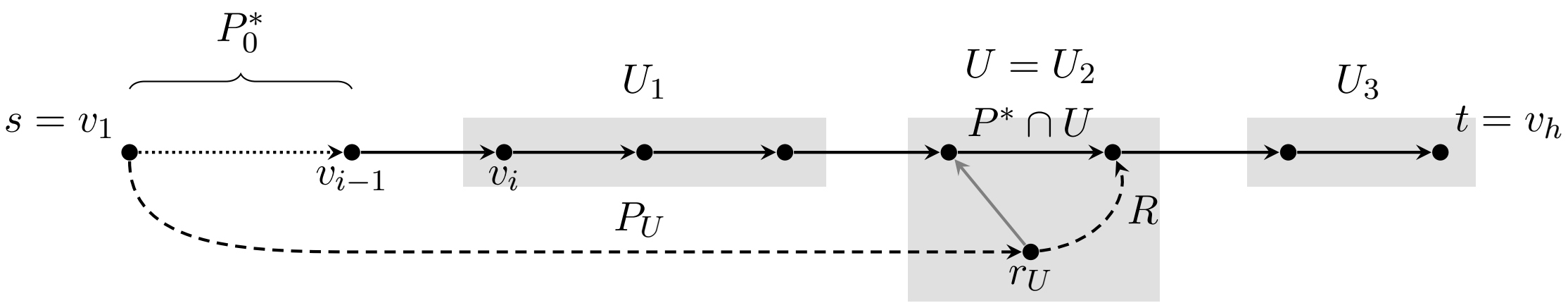

If $f(P^*\cap U\mid P_U)=f(P^*\cap U)$, conditioning loses none of the value
of $P^*\cap U$. The stored path leaves at most $\lambda\OPT/\alpha$ to reach
the target, whereas $f(P^*\cap U)/\alpha'\ge \lambda\OPT/\alpha$ by
\eqref{eq:overview-segment} and \eqref{eq:overview-rate}. Hence the combined
path has value at least $\OPT/\alpha$. The full analysis handles intermediate
cases with the same loss of $O(\log^2 n)$ per decomposition level.
Figure~\ref{fig:optimal-path} illustrates the strategy.

Let $d$ be the maximum number of nested suffix recursions. Starting from
a factor of order $1/\lambda$, iterating \eqref{eq:overview-rate} gives
approximation rate $\lambda^{-1}O(\log^2 n)^d$. In the simplified picture,
each suffix recursion reduces the optimum by a factor at most $\lambda$, and we can
stop once the subproblem optimum is at most $\OPT/n^2$. Thus
\[
  d\le\left\lceil\log_{1/\lambda}(n^2)\right\rceil
  =O\!\left(\sqrt{\frac{\log n}{\log\log n}}\right).
\]
This gives the claimed subpolynomial approximation rate
$2^{O(\sqrt{\log n\,\log\log n})}$.

For the running time, first bound the depth. In the same simplified picture,
every step multiplies the optimum by at most $1-\lambda$ (which happens in prefix recursions), and recursion stops
at $\OPT/n^2$. The depth is therefore $O(\log n/\lambda)$.

Next fix a vertex and count calls containing it at one level. Computing paths
to all destinations together allows one prefix recursion and at most one
suffix recursion per part. The vertex therefore belongs to at most one child of each
type. Their optima sum to at most the parent's optimum, so the sum at any
level is at most the initial optimum. Every non-leaf call has optimum greater
than $\OPT/n^2$, giving fewer than $n^2$ such calls per level. Multiplying by
the depth and the $n$ choices of the vertex gives $O(n^3\log n/\lambda)$ non-leaf
calls. Section~\ref{sec:runtime} gives the
formal running-time argument.

\subsection{Formal definition of the algorithm}

The procedure $\ADSO(G,s,f,\tau)$ seeks an $s$--$v$ path of maximal value for each vertex $v$, in the form of a table of paths $D[v]$, $v\in V$.
For a given destination no further improvement is needed
once the stored value reaches \emph{threshold} $\tau$.
Write $\OPT(v)$ for the maximum value of
an $s$--$v$ path, which we assume exists for each vertex $v$. Fix a rational parameter $0<\lambda\le1/2$ for the entire
execution. It is implicitly passed to recursive calls without changes.
At the top level, run $\ADSO(G,s,f,f(V))$ and output the path
returned for $t$. Section~\ref{sec:guarantee} proves the guarantee stated in
Algorithm~\ref{alg:adso}.

\begin{algorithm}{All-Destinations Submodular Orienteering}
  \label{alg:adso}
  \textbf{Input:} Transitively closed DAG $G=(V,A)$, source $s$ reaching every vertex,
  normalized monotone submodular $f$, threshold $\tau>0$.\par
  \textbf{Output:} An $s$--$v$ path $D[v]$ for every $v\in V$.
  For the approximation rate $\alpha$ established in
  Section~\ref{sec:guarantee}, the returned paths satisfy
  \[
    f(D[v])\ge\min\left\{\tau,\frac{\OPT(v)}{\alpha}\right\}
    \qquad\forall v\in V.
  \]
  \begin{tabbing}
    \hspace{1.2em}\=\hspace{1.3em}\=\hspace{1.3em}\=\kill
    \>\emph{Base case}\\
    1\>Set $D[s]\leftarrow(s)$ and $D[v]\leftarrow(s,v)$ for $v\ne s$; propagate with $f$\\
    2\>\textbf{if} $f(D[v])\in\{0\}\cup[\tau,\infty)$ for every $v$, \textbf{return} $D$\\
    \>\emph{Main algorithm}\\
    3\>$D\leftarrow\ADSO(G,s,f,(1-\lambda)\tau)$\\
    4\>$V_+\leftarrow\{v:f(D[v])\ge(1-\lambda)\tau\}$\\
    5\>Construct $\mathcal P$ using Lemma~\ref{lem:partition} on $G[V_+]$\\
    6\>\textbf{for} $U\in\mathcal P$ with $f(D[r_U])<\tau$ \textbf{do}\\
    7\>\>$\tau'\leftarrow\tau-f(D[r_U])$\\
    8\>\>$D'\leftarrow\ADSO(G[U],r_U,f(\,\cdot\mid D[r_U]),\tau')$\\
    9\>\>\textbf{for} $u\in U$ with $f(D[r_U]\circ D'[u])>f(D[u])$ \textbf{do}\\
    10\>\>\>$D[u]\leftarrow D[r_U]\circ D'[u]$\\
    11\>Propagate $D$ with $f$ in $G$ and \textbf{return} $D$
  \end{tabbing}
\end{algorithm}

The \emph{propagate} operation on table $D$ tries to make straightforward improvements to the paths in $D$.
It visits each vertex $u$ in
topological order and inspects every outgoing arc $(u,v)$. It performs the
update
\[
  D[v]\leftarrow D[u]\circ(v)
  \quad\text{if}\quad
  f(D[u]\circ(v))>f(D[v]).
\]
When $u$ is processed, all updates to $D[u]$ are complete. Since appending $v$
cannot decrease reward, the final table satisfies
\begin{equation}
  \label{eq:monotone}
  f(D[u])\le f(D[v])\qquad\forall (u,v)\in A.
\end{equation}
Starting from $D[s]=(s)$ and $D[v]=(s,v)$ for $v\ne s$, propagation also
ensures that $f(D[v])\ge f(\{u\})$ for every vertex $u$ that can reach $v$.
Consequently, $f(D[v])=0$ certifies that every path from $s$ to $v$ has value
zero, by subadditivity. If every stored value is either zero or at least
$\tau$, the table already satisfies the output guarantee stated in Algorithm~\ref{alg:adso}. This is the
base case of the recursion.

The root $r_U$ of a part $U$ reaches every vertex in $G[U]$, so it is a valid
source for the suffix recursion. Any path $Q$ returned by this call starts at $r_U$
and can be concatenated to $D[r_U]$. A repeated vertex other than $r_U$ would
create a directed cycle, so the concatenation is a path and
\begin{equation}
  \label{eq:join}
  f(D[r_U]\circ Q)=f(D[r_U])+f(Q\mid D[r_U]).
\end{equation}
This intuitively motivates why we can use the same marginal objective for the
entire part.

If $V_+=\emptyset$, use the empty partition. Each entry $D[r_U]$ stays fixed
throughout the loop: other parts do not update it, and the only $r_U$--$r_U$
path is $(r_U)$, so its own update cannot improve it. If $f(D[r_U])\ge\tau$,
every vertex in that part already has stored value at least $\tau$, so no
further call is needed. Otherwise, its marginal threshold satisfies
\[
  0<\tau'=\tau-f(D[r_U])\le\lambda\tau.
\]
\section{Approximation guarantee}
\label{sec:guarantee}

We make the overview precise by following the thresholds in the recursion. In
this section and Section~\ref{sec:runtime}, let $n$ denote the number of
vertices in the initial instance and $\tau_0=f(V)>0$ its initial threshold. In
a recursive call, $f$ and $\OPT(v)$ refer to that call's objective and
optimum. The prefix recursion uses the same graph, source, and objective with
threshold $(1-\lambda)\tau$. Each suffix recursion uses a rooted subgraph, its root
as the source, and a marginal objective with threshold at most $\lambda\tau$.

Every recursive objective is a marginal of the initial one. The marginal
$f(\{v\}\mid S)$ is zero if $v$ belongs to $S$ or to a set that a parent call
conditioned on; otherwise, the assumption in Lemma~\ref{lem:preprocessing} gives a marginal
of at least $\tau_0/n^2$. Thus every positive objective value is at least
$\tau_0/n^2$, since it is a sum of marginal gains. Consequently, every call
with threshold at most $\tau_0/n^2$ is a base case; the test in line~2 of
Algorithm~\ref{alg:adso} detects this automatically. Each step reduces the
threshold by a factor at most $1-\lambda<1$, so the recursion terminates.
Along any root-to-leaf path, the number of suffix recursions is at most
\begin{equation}
  \label{eq:depth}
  d_{\max}=\left\lceil\log_{1/\lambda}(n^2)\right\rceil.
\end{equation}
As in the overview, we lose a factor $O(\log^2 n)$ for each nested suffix recursion.
Choose a constant $c$
such that $c\lceil\log n\rceil^2$ is at least twice the path-intersection
bound in Lemma~\ref{lem:partition} for every subproblem of the initial
instance. For every nonnegative integer $d$, set
\begin{equation}
  \label{eq:factor}
  \alpha_d=\frac{\bigl(c\lceil\log n\rceil^2\bigr)^d}{\lambda}.
\end{equation}

\subsection{The recursive guarantee}

As in the overview, the prefix recursion concerns a prefix of an optimal path, and
the suffix recursions concern the intersections with the rooted subgraphs.

\begin{lemma}
  \label{lem:threshold}
  Let $D=\ADSO(G,s,f,\tau)$, and let $d\ge0$ be an integer bounding the number
  of suffix recursions on every root-to-leaf path below this call. For every
  endpoint $v$,
  \begin{equation}
    \label{eq:threshold-guarantee}
    f(D[v])<\tau
    \quad\Longrightarrow\quad
    \OPT(v)\le\alpha_d\tau.
  \end{equation}
\end{lemma}

\begin{proof}
  We perform an induction over the recursion tree, proving the claim for every
  $d$. Fix an endpoint $t$ with $f(D[t])<\tau$. In a base case, the stored
  value at $t$ is zero, so every $s$--$t$ path has value zero and the claim
  holds. Otherwise, fix an optimal $s$--$t$ path $P^*$ and let $D_0$ be the
  table returned by the prefix recursion. This call has threshold $(1-\lambda)\tau$
  and satisfies the induction hypothesis with the same $d$. If $t\notin V_+$,
  it already gives $\OPT(t)\le\alpha_d(1-\lambda)\tau \le\alpha_d \tau$.

  Assume now that $t\in V_+$. By \eqref{eq:monotone}, no arc leaves $V_+$, so
  the vertices of $P^*$ outside $V_+$ form a prefix $P_0^*$. The remaining
  suffix intersects $q=O(\log^2 n)$ parts. Every nonempty $P^*\cap U$ is a
  path in $G[U]$ by transitive closure. All sums below range over these
  intersected parts.

  \smallskip\noindent
  \emph{Prefix recursion.}
  If $P_0^*$ is nonempty, the prefix recursion returns value below
  $(1-\lambda)\tau$ at its last vertex. Its induction bound gives
  \begin{equation}
    \label{eq:low}
    f(P_0^*)\le\alpha_d(1-\lambda)\tau.
  \end{equation}
  The same inequality holds if the prefix is empty, since $f(\emptyset)=0$.

  \smallskip\noindent
  \emph{Suffix recursion.}
  Fix an intersected part $U$ and let $P_U=D_0[r_U]$ be the stored path to its
  root, as in the overview. Let $u$ be the last vertex of $P^*\cap U$. We have
  \[
    (1-\lambda)\tau\le f(P_U)<\tau.
  \]
  Indeed, the lower bound follows from $r_U\in V_+$; if $f(P_U)\ge\tau$,
  propagation through $u$ to $t$ would contradict $f(D[t])<\tau$. Thus the
  suffix recursion on $U$ has threshold $\tau'=\tau-f(P_U)\le\lambda\tau$. In
  particular, $d\ge1$, and the induction hypothesis applies to this child with
  $d-1$.

  Let $R$ be the path returned by this call for $u$. The algorithm considers
  $P_U\circ R$ and propagates its value to $t$. Hence $f(R\mid P_U)<\tau'$.
  Prepending $r_U$ to $P^*\cap U$ when necessary gives a feasible $r_U$--$u$
  path without changing its marginal value at $P_U$. Induction therefore gives
  \begin{equation}
    \label{eq:high}
    f(P^*\cap U\mid P_U)\le\alpha_{d-1}\bigl(\tau-f(P_U)\bigr).
  \end{equation}
  This bounds the marginal value not already covered by $P_U$. By monotonicity
  and \eqref{eq:high},
  \[
    \begin{aligned}
      f(P^*\cap U)
      &\le f(P_U)+\alpha_{d-1}\bigl(\tau-f(P_U)\bigr)\\
      &\le\bigl(1-\lambda+\lambda\alpha_{d-1}\bigr)\tau\\
      &\le2\lambda\alpha_{d-1}\tau,
    \end{aligned}
  \]
  where we use $f(P_U)\ge(1-\lambda)\tau$ and $\alpha_{d-1}\ge1/\lambda$.

  \smallskip\noindent
  \emph{Combining the pieces.}
  Subadditivity and the bounds above give
  \[
    \begin{aligned}
      \OPT(t)
      &\le f(P_0^*)+\sum_U f(P^*\cap U)\\
      &\le\bigl((1-\lambda)\alpha_d+2q\lambda\alpha_{d-1}\bigr)\tau\\
      &\le\alpha_d\tau.
    \end{aligned}
  \]
  The last inequality follows from $2q\alpha_{d-1}\le\alpha_d$, by the choice
  of the constant in \eqref{eq:factor}.
\end{proof}

\subsection{The output guarantee}

Following only prefix recursions gives progressively smaller thresholds for the
same graph, source, and objective. Their paths are retained or improved by
every caller. This yields the approximation guarantee stated in
Algorithm~\ref{alg:adso}.

\begin{corollary}
  \label{cor:approximation}
  Let $D=\ADSO(G,s,f,\tau)$, and let $d\ge0$ be an integer bounding the number
  of nested suffix recursions below this call. For every endpoint $v$,
  \[
    f(D[v])\ge\min\left\{\tau,\frac{(1-\lambda)\OPT(v)}{\alpha_d}\right\}.
  \]
\end{corollary}

\begin{proof}
  Fix $v$ and write $a=f(D[v])$. If $a\ge\tau$ or $\OPT(v)=0$, the claim is
  immediate. Otherwise, follow the chain of prefix recursions. Since $\OPT(v)>0$,
  propagation gives positive value at $v$ in every call on this chain. These
  returned values are at most $a$. The last call is a base case, so its
  threshold is at most $a$. Since the first threshold is $\tau>a$, the chain
  contains consecutive thresholds $\rho$ and $(1-\lambda)\rho$ with
  \[
    (1-\lambda)\rho\le a<\rho.
  \]
  The call at threshold $\rho$ returns value below $\rho$ at $v$ and has at
  most $d$ nested suffix recursions below it. Lemma~\ref{lem:threshold} therefore
  gives
  \[
    \OPT(v)\le\alpha_d\rho\le\frac{\alpha_d a}{1-\lambda}.\qedhere
  \]
\end{proof}

The output guarantee in Algorithm~\ref{alg:adso} thus holds with
$\alpha=\alpha_d/(1-\lambda)$. At the top level, $\tau_0=f(V)\ge\OPT(v)$ for
every $v$, so $f(D[v])\ge\OPT(v)/\alpha$. Taking $d=d_{\max}$ from
\eqref{eq:depth} and returning $D[t]$ gives approximation rate
$\alpha_{d_{\max}}/(1-\lambda)$.

\section{Running time and proof of the main theorem}
\label{sec:runtime}

We follow the two steps of the overview: bound the depth of the recursion
tree, then count calls containing a fixed vertex at each level. The quantities
that decrease are the thresholds. Recall that a child from a prefix recursion has threshold
$(1-\lambda)\tau$, a child from a suffix recursion has threshold at most $\lambda\tau$, and
every call with threshold at most $\tau_0/n^2$ is a leaf.

\paragraph{Depth of the recursion tree.}
Since $\lambda\le1-\lambda$, a call at depth $h$ has threshold at most
$\tau_0(1-\lambda)^h\le\tau_0e^{-\lambda h}$. Every call at depth
\[
  h_{\max}=\left\lceil\frac{2\ln n}{\lambda}\right\rceil
  =O(\log n/\lambda)
\]
is therefore a leaf.

\paragraph{Number of recursive calls.}
Fix an original vertex $v$ and consider only calls containing it. These form a
subtree rooted at the initial call. Each node has at most one child from a prefix recursion
and one from a suffix recursion, because the suffix recursion subgraphs are disjoint. The thresholds
of these children sum to at most
\[
  (1-\lambda)\tau+\lambda\tau=\tau.
\]
By induction, the sum of thresholds at any level of this subtree is at most
$\tau_0$. Every non-leaf call has threshold greater than $\tau_0/n^2$, so
there are fewer than $n^2$ non-leaf calls per level and at most $n^2h_{\max}$
altogether. Each has at most two children in this subtree, which therefore
contains at most $1+2n^2h_{\max}$ calls, including leaves. Every call contains
an original vertex. Summing over the $n$ choices of $v$ gives
\begin{equation}
  \label{eq:runtime}
  N_{\mathrm{calls}}\le2n^3h_{\max}+n
  =O(n^3\log n/\lambda).
\end{equation}

Each call performs polynomial work.

\paragraph{Choice of parameters.}
Write $x=\log(1/\lambda)$. With $d_{\max}=\lceil2\log n/x\rceil$,
\eqref{eq:factor} gives
\[
  \log\alpha_{d_{\max}}=x+O(d_{\max}\log\log(2n))
  =O\!\left(x+\frac{\log n\,\log\log(2n)}{x}\right).
\]
Balancing the two terms gives $x=\Theta(\sqrt{\log n\,\log\log n})$, as in the
overview. For $n\ge2$, use the following dyadic choice:
\begin{equation}
  \label{eq:parameters}
  \begin{aligned}
    k&=\left\lceil\sqrt{\log(2n)\,\log\log(4n)}\right\rceil,
    &\lambda&=2^{-k},
    &d_{\max}&=\left\lceil\frac{2\log n}{k}\right\rceil.
  \end{aligned}
\end{equation}
Then $n^2\lambda^{d_{\max}}\le1$ and $\lambda\le1/2$. By \eqref{eq:runtime},
\[
  N_{\mathrm{calls}}=O(n^3 2^k\log n)=n^{3+o(1)}.
\]
Each call takes polynomial time, so the running time is polynomial.

Corollary~\ref{cor:approximation} and the preprocessing in Lemma~\ref{lem:preprocessing} give approximation
rate at most $6\alpha_{d_{\max}} / (1 - \lambda) \le 12 \alpha_{d_{\max}}$, and
\[
	\log(12\alpha_{d_{\max}})
  =O\!\left(\log(1/\lambda)+d_{\max}\log\log(2n)\right)
  =O\bigl(\sqrt{\log n\,\log\log n}\bigr).
\]
This proves Theorem~\ref{thm:main} assuming Lemmas~\ref{lem:partition} and~\ref{lem:preprocessing}.

\section{Instance simplification}
\label{sec:length}

In this section we prove Lemma~\ref{lem:preprocessing}, which we restate here for convenience:
\begin{samepage}
\preprocessing*
\end{samepage}

Prior to our work, it was already known that unbounded instances are equivalent 
up to constant factors to instances with
polynomially bounded integer lengths and budgets, see for example~\cite[Section~5]{CRSVZ26}.
We will use this within our reduction.
All properties not related to the length bound are simple observations that also appear
in various previous works.
We provide a self-contained proof of all properties needed for the lemma.

\subsection{Basic reductions}
\label{sec:basic-reductions}

\paragraph{From walks to a DAG.}
Compute shortest-path distances in the original $n$-vertex graph, with zero
distance from each vertex to itself. The DAG has vertices $(v,i)$,
$i=0,\ldots,n$, and an arc from $(u,i)$ to $(v,i+1)$ of the shortest $u$--$v$
distance whenever it is finite and $i<n$. The start and end vertices are
$(s,0)$ and $(t,n)$. Evaluate rewards by applying $f$ to the distinct original
vertices represented; this remains monotone and submodular.

Consider now an $s$--$t$ walk in the original graph and list its distinct
vertices $v_1,\ldots,v_h$ in order of first appearance, where $v_1=s$ and
$h\le n$. This induces the path
\[
  (s=v_1,0),(v_2,1),\ldots,(v_h,h-1),
  (t,h),\ldots,(t,n)
\]
in the DAG. The shortest-path arcs and zero-length padding give no greater
length and the same reward. Conversely, for a path in the DAG, we can expand
each arc into a shortest path, preserving length and not decreasing reward.
Thus the optimum is preserved, with $O(n^2)$ vertices.

\paragraph{Polynomially bounded budgets.}
Consider an $n$-vertex DAG with nonnegative integer lengths and a polynomially
bounded integer budget $B$. Create a graph with states $(v,b)$,
$b=0,\ldots,B$, as vertices; $b$ records the distance travelled. For each arc
$(u,v)$, add an arc from $(u,b)$ to $(v,b+\ell(u,v))$ whenever
$b+\ell(u,v)\le B$. The starting vertex is $(s,0)$ and the end vertex is
$(t,b)$ for each choice $b\in\{0,\ldots,B\}$; discard infeasible choices and
keep the best path. Evaluate rewards on represented vertices as in the
previous reduction. Ordering states by an original topological order proves
acyclicity, even with zero lengths. Its $(s,0)$--$(t,b)$ paths correspond
exactly to original $s$--$t$ paths of length $b$, with the same reward, and it
has $n(B+1)$ vertices. Conversely, an unbounded DAG is a budgeted instance
with zero lengths and budget. Thus the equivalence preserves the optimum with
a polynomial increase in size. At the unbounded stage, take the transitive
closure before invoking $\ADSO$. Expanding its added arcs back into paths
preserves feasibility and can only increase reward, so this also preserves the
optimum.

\paragraph{Enforcing the marginal bound.}
Consider an arbitrary objective $f$ with $n\ge2$ and $f(V)>0$; the other cases
are immediate. Replace $f$ by
\[
  \widehat f(S)=f(S)+\frac{f(V)}{n(n-1)}|S|.
\]
This is normalized, monotone, and submodular. For $v\notin S$, its marginal
value satisfies
\[
  \widehat f(\{v\}\mid S)
  \ge\frac{f(V)}{n(n-1)}
  =\frac{\widehat f(V)}{n^2}.
\]
Thus $\widehat f$ satisfies the bound in Lemma~\ref{lem:preprocessing}.

Let $M=\max_{v\in V}f(\{v\})$. Since every remaining vertex lies on an
$s$--$t$ path, we can keep as a candidate such a path $Q$ through a vertex of
singleton value $M$. For every path $P$, the added reward is at most
$f(V)/(n-1)\le nM/(n-1)\le2M$. Write $\OPT_f$ and $\OPT_{\widehat f}$ for the
optimal path values under the two objectives. Suppose the algorithm returns a
path $P$ with approximation rate $\gamma\ge1$ for $\widehat f$, so
$\widehat f(P)\ge\OPT_{\widehat f}/\gamma$. Since $f(Q)\ge M$, returning the
better of $P$ and $Q$ according to $f$ gives value at least
\[
  \max\{f(P),f(Q)\}
  \ge\frac{f(P)+2M}{3}
  \ge\frac{\widehat f(P)}{3}
  \ge\frac{\OPT_f}{3\gamma}.
\]
The preprocessing therefore loses at most a factor three.

\subsection{Repairing a budget violation}
\label{sec:repairing}

The remaining construction applies to a feasible length-bounded instance on a
DAG, with budget $C$. Here $n$ denotes the number of vertices in this DAG,
after the walk reduction if it was needed. Delete vertices on no $s$--$t$
path. Take the transitive closure and assign each arc $(u,v)$ the length of a
shortest $u$--$v$ path, continuing to write $\ell(u,v)$ for its length. This
preserves the optimum: every original feasible path remains feasible, and
expanding closure arcs gives an original path of the same length and at least
the same reward. The lengths satisfy the triangle inequality
\[
  \ell(u,v)\le \ell(u,w)+\ell(w,v)
\]
whenever $(u,w)$ and $(w,v)$ are arcs. To simplify notation, set $\ell(v,v)=0$
without adding self-loops. Define the slack of a vertex $v$ by
\[
  S(v)=C-\ell(s,v)-\ell(v,t).
\]
A vertex of negative slack cannot lie on a feasible path, so delete these
vertices. The lengths $\ell$ remain fixed throughout the construction.

\begin{lemma}
  \label{lem:repair}
  Let $P$ be an $s$--$t$ path of length $C+\eta$, with $\eta>0$. If every
  vertex of $P$ has slack at least $\eta$, there are two budget-feasible
  $s$--$t$ paths that cover $V(P)$. They can be found in polynomial time, and
  one has reward at least $f(P)/2$.
\end{lemma}

\begin{figure}
  \centering
  \begin{tikzpicture}[x=1.25cm,y=0.7cm,>=stealth,
      every node/.style={font=\normalsize}]
    \begin{scope}[yshift=-12mm]
      \foreach \x/\name in {0/bs,4.2/bu,5.8/bv,10/bt}
      \node[circle,fill=black,inner sep=1.55pt] (\name) at (\x,-3.3) {};
      \draw[->,very thick] (bs)--(bu);
      \draw[->,very thick, black!50] (bu)--(bv);
      \draw[->,very thick] (bv)--(bt);
      \draw[->,thick,densely dashed] (bs)
        .. controls (1.5,-1.9) and (4.3,-1.9) .. (bv);
      \draw[->,thick,densely dashed] (bu)
        .. controls (5.7,-4.8) and (8.5,-4.8) .. (bt);
      \node[above] at (2.7,-2.05) {arc $(s,v)$};
      \node[below] at (7.3,-4.75) {arc $(u,t)$};
      \node[below left] at (bs) {$s$};
      \node[above] at (bu) {$u$};
      \node[below] at (bv) {$v$};
      \node[above right] at (bt) {$t$};
      \node at (2.1,-3.65) {prefix of $P$};
      \node at (7.9,-2.95) {suffix of $P$};
    \end{scope}
  \end{tikzpicture}
  \caption{Repairing a path of length $C+\eta$}

  \label{fig:length-reduction}
\end{figure}
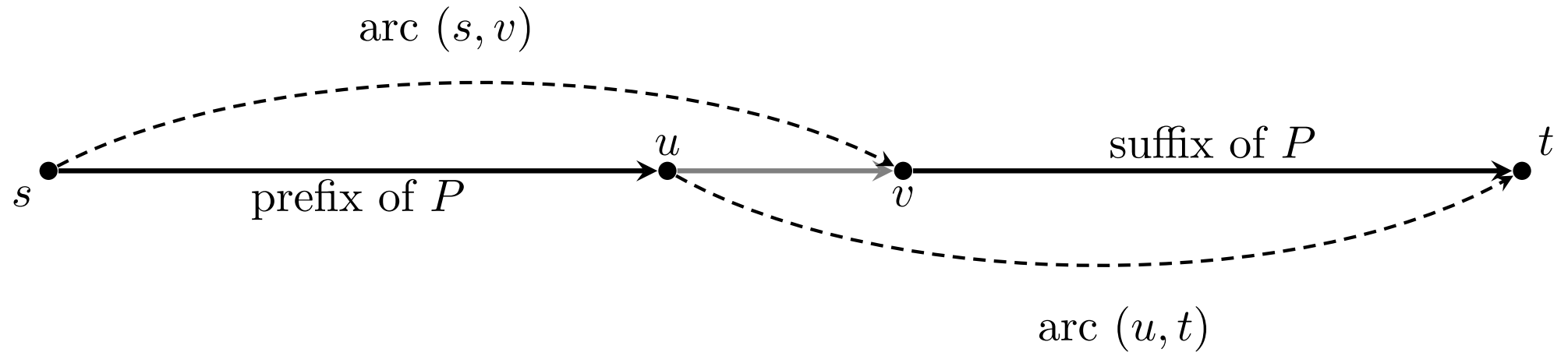

\begin{proof}
  Traverse $P$ from $s$, and let $v$ be the first vertex whose prefix has
  length at least $\ell(s,v)+\eta$. Such a vertex exists, because $t$ satisfies it.
	Further, it cannot be $s$. Replacing this prefix by the arc $(s,v)$ saves at least $\eta$, giving
  a budget-feasible path.

  Let $u$ immediately precede $v$ on $P$. Its prefix has length less than
  $\ell(s,u)+\eta$. Appending the arc $(u,t)$ therefore gives length less than
  \[
    \ell(s,u)+\eta+\ell(u,t)\le C.
  \]
  Both concatenations follow the vertex order of $P$, so they are paths. Their
  vertex sets have union $V(P)$, since $u$ and $v$ are consecutive; see
  Figure~\ref{fig:length-reduction}. Subadditivity gives the reward bound.
\end{proof}

\subsection{Rounding the remaining budget}
\label{sec:rounding}

We guess a vertex $w$ of minimum slack on an optimal path by trying every
vertex. Each guess gives a separate instance; we return the best feasible path
obtained. Fix a guess $w$, set $\sigma=S(w)$, and retain only vertices $v$
with $S(v)\ge\sigma$. This preserves every path on which $w$ has minimum
slack. To restrict to paths through $w$, take the union of the subgraphs
induced by its ancestors and descendants in the retained graph, including $w$
in both. They intersect only at $w$, and we keep only arcs within one of these
subgraphs, so every surviving $s$--$t$ path passes through $w$. Let $A_-$ and
$A_+$ be the arc sets of these two subgraphs, respectively. Give each arc
$(u,v)$ the nonnegative reduced length
\[
  \ell'(u,v)=
  \begin{cases}
    \ell(u,v)+\ell(s,u)-\ell(s,v),&(u,v)\in A_-,\\
    \ell(u,v)+\ell(v,t)-\ell(u,t),&(u,v)\in A_+.
  \end{cases}
\]
Nonnegativity follows from the triangle inequality. For an $s$--$t$ path $P$
through $w$ with arc set $A(P)$,
\begin{equation}
  \label{eq:remaining-budget}
  \begin{aligned}
    \ell'(P)
    &=\sum_{(u,v)\in A(P)\cap A_-}\bigl(\ell(u,v)+\ell(s,u)-\ell(s,v)\bigr)\\
    &\quad+\sum_{(u,v)\in A(P)\cap A_+}\bigl(\ell(u,v)+\ell(v,t)-\ell(u,t)\bigr)\\
    &=\ell(P)-\ell(s,w)-\ell(w,t)=\ell(P)-C+\sigma.
  \end{aligned}
\end{equation}
Thus every budget-feasible path through $w$ has reduced length at most
$\sigma$.

If $\sigma=0$, retain only zero-reduced-length arcs; the instance is already
unbounded, and every surviving path has length $C$ by
\eqref{eq:remaining-budget}.

If $\sigma>0$, set the rounded length of each arc to
\[
  \ell''(u,v)=\left\lfloor\frac{n\ell'(u,v)}\sigma\right\rfloor,
\]
discard arcs whose rounded length exceeds $n$, and set the budget to $n$. This
gives a polynomially bounded instance on the same vertices with the same
reward function $f$.

Every feasible path $P$ through $w$ survives the rounding, since
$\ell''(P)\le n\ell'(P)/\sigma\le n$. For each arc, rounding loses less than
$\sigma/n$ in reduced length:
\[
  0\le\ell'(u,v)-\frac\sigma n\ell''(u,v)<\frac\sigma n.
\]
Conversely, a path $P$ feasible for the rounded instance has $\ell''(P)\le n$
and $|A(P)|\le n-1$. Summing the rounding errors over its arcs gives
\[
  0\le\ell'(P)-\frac\sigma n\ell''(P)
  \le\frac{|A(P)|\sigma}{n}
  \le\frac{(n-1)\sigma}{n}<\sigma.
\]
Consequently, $\ell'(P)<\sigma+\frac\sigma n\ell''(P)
\le2\sigma$.
By \eqref{eq:remaining-budget}, $\ell(P)<C+\sigma$. Since every vertex has
slack at least $\sigma$, Lemma~\ref{lem:repair} repairs any budget violation
while retaining at least half the reward.

\subsection{Proof of Lemma~\ref{lem:preprocessing}}
We first reduce the given instance to a DAG instance as in Section~\ref{sec:basic-reductions}.
Then we perform the transformations in Sections~\ref{sec:repairing} and~\ref{sec:rounding},
which create one instance for each guess of the minimum slack vertex $w$.
Reduce each resulting instance to an unbounded one as in Section~\ref{sec:basic-reductions}.
Remove all vertices that are not on $s$--$t$ paths, and replace the graph by its transitive closure.
Next, modify the objective function to guarantee the bounds on the marginals, as described also in
Section~\ref{sec:basic-reductions}.
For an $\alpha$-approximation of each of the resulting instances,
replacing the objective function again by the original one and choosing the better of 
$P$ and $Q$ (as in Section~\ref{sec:basic-reductions})
loses a factor of at most three.
We then undo the
unbounded reduction. For a guess $w$ of minimum slack on an optimal path, that path
remains feasible, so the rounded optimum is at least the original optimum.
The repair loses at most a factor two and
choosing the best feasible solution over all
guesses preserves this guarantee.

Finally, expand the weighted closure arcs and undo the walk reduction. These
expansions preserve length and can only increase reward. This
completes the proof of Lemma~\ref{lem:preprocessing}.

\section{Deterministic decomposition}
\label{sec:decomposition}

We will now prove the deterministic decomposition lemma, which we restate here for convenience:
\begin{samepage}
\partition*
\end{samepage}
The randomized construction in Section~\ref{sec:algorithm} groups vertices by
their first ancestor in a random ordering. We keep this construction but
choose the ordering deterministically. The following lemma states the
auxiliary properties that we will enforce, which in turn imply the properties
of Lemma~\ref{lem:partition}, as we show afterwards.

\begin{lemma}
  \label{lem:ordering}
  Let $G=(V,A)$ be a directed graph, and let $\Anc(v)$ consist of $v$ and all
  vertices that can reach it. Set $k=\lceil\log(2n^2)\rceil$. In deterministic
  polynomial time, we can order $V$ so that, for every $u,v\in V$ with $u$
  reaching $v$ and $k\le|\Anc(v)|\le2|\Anc(u)|$, the first $k$ vertices of
  $\Anc(v)$ contain at least one vertex of $\Anc(u)$.
\end{lemma}

\begin{figure}[ht]
  \centering
  \begin{tikzpicture}[x=1cm,y=1cm,>=stealth,
      every node/.style={font=\normalsize}]
    \fill[black!5] (0,1.25)
      .. controls (3.2,1.55) and (9,1.15) .. (9,0)
      .. controls (9,-1.15) and (3.2,-1.55) .. (0,-1.25)--cycle;
    \fill[black!13] (0,0.75)
      .. controls (1.8,0.95) and (4.8,0.75) .. (4.8,0)
      .. controls (4.8,-0.75) and (1.8,-0.95) .. (0,-0.75)--cycle;
    \draw[thick,black!60] (0,1.25)
      .. controls (3.2,1.55) and (9,1.15) .. (9,0)
      .. controls (9,-1.15) and (3.2,-1.55) .. (0,-1.25);
    \draw[thick,black!75] (0,0.75)
      .. controls (1.8,0.95) and (4.8,0.75) .. (4.8,0)
      .. controls (4.8,-0.75) and (1.8,-0.95) .. (0,-0.75);
    \fill[white,path fading=east,overlay] (-0.1,-1.65) rectangle (1.25,1.65);
    \node at (1.6,0) {$\Anc(u)$};
    \node at (6.65,-0.58) {$\Anc(v)$};
    \node[circle,fill=black,inner sep=1.55pt] (u) at (4.8,0) {};
    \node[circle,fill=black,inner sep=1.55pt] (v) at (9,0) {};
    \draw[->,thick] (u)
      .. controls (5.3,0.3) and (5.6,0.3) .. (6,0)
      .. controls (6.4,-0.3) and (6.6,0.22) .. (7,0.07)
      .. controls (7.6,-0.17) and (8.2,0) .. (v);
    \node[below=4pt] at (u) {$u$};
    \node[below=4pt] at (v) {$v$};
	  \draw[densely dashed] (4.8, 0.75) ellipse (2cm and 0.4cm);
    \node at (4.8,0.75) {first $k$};
  \end{tikzpicture}
  \caption{$u$ and $v$ as in Lemma~\ref{lem:ordering}}

  \label{fig:ancestor-sets}
\end{figure}
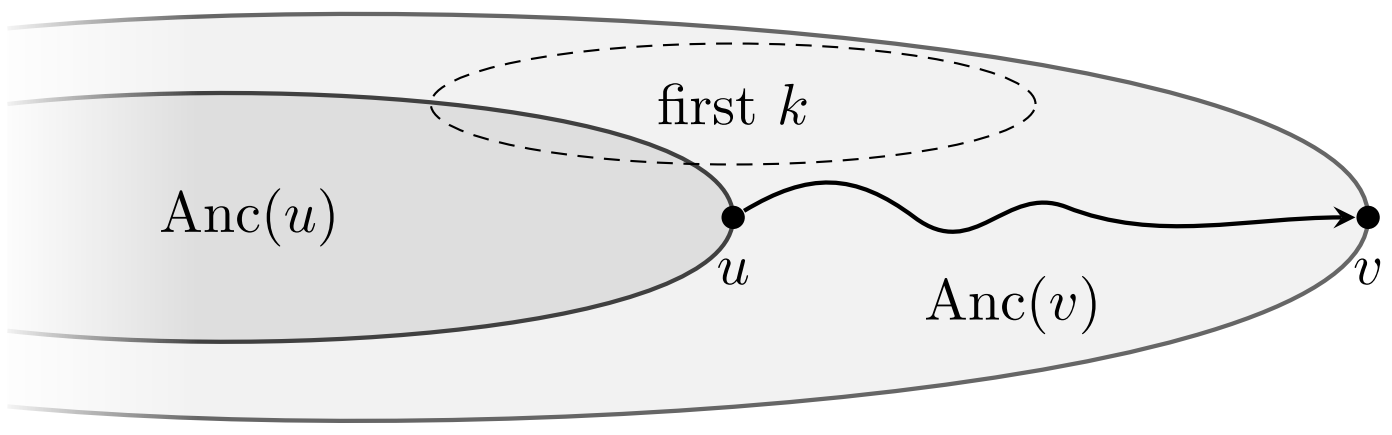

\begin{proof}
  For a uniformly random ordering of $V$, call a pair $(u,v)$ bad if the
  stated property fails. Since $\Anc(u)\subseteq\Anc(v)$, sampling without
  replacement gives
  \[
    \mathbb P[(u,v)\text{ is bad}]
    \le\left(1-\frac{|\Anc(u)|}{|\Anc(v)|}\right)^k\le2^{-k}.
  \]
  There are at most $n^2$ pairs, so the expected number of bad pairs is at
  most $1/2$.

  We fix the ordering one vertex at a time, keeping this conditional
  expectation at most $1/2$. Try each remaining vertex as the next vertex and
  choose one that does not increase the conditional expectation. Such a choice
  exists because averaging over the next vertex gives the current conditional
  expectation. At the end, the number of bad pairs is an integer at most
  $1/2$, hence zero.

  The only task that remains is to efficiently compute the conditional
  expectation for each already determined prefix. After fixing a prefix, an
  already decided pair $(u,v)$ contributes zero or one, which can easily be
  determined. For an undecided pair, suppose $j<k$ vertices of $\Anc(v)$ and
  no vertex of $\Anc(u)$ have appeared. The conditional probability of a bad
  pair is
  \[
    \frac{\binom{|\Anc(v)\setminus\Anc(u)|-j}{k-j}}
    {\binom{|\Anc(v)|-j}{k-j}},
  \]
  where the numerator is zero if too few vertices remain in
  $\Anc(v)\setminus\Anc(u)$. These probabilities are computable exactly with
  polynomial encoding length. There are polynomially many steps, candidates,
  and pairs.
\end{proof}

\noindent\begin{minipage}{\linewidth}
  \setlength{\parindent}{17pt}
  \begin{proof}[Proof of Lemma~\ref{lem:partition}]
    Set $k=\lceil\log(2n^2)\rceil$. Use the ordering constructed in
    Lemma~\ref{lem:ordering}. Group vertices by their first ancestor in this
    ordering, as in the randomized construction. As also explained in
    Section~\ref{sec:algorithm}, each resulting subgraph is rooted for any
    given ordering.

    Fix any directed path $P=(v_1,\ldots,v_h)$. Its ancestor sets satisfy
    $\Anc(v_1)\subseteq\cdots\subseteq\Anc(v_h)$. Divide $P$ into consecutive
    segments according to $\lfloor\log|\Anc(v_i)|\rfloor$; there are at most
    $1+\lfloor\log n\rfloor$ segments. Let $v_i$ and $v_j$ be the first and
    last vertices of one segment. Then $\Anc(v_i)\subseteq\Anc(v_j)$ and
    $|\Anc(v_j)|<2|\Anc(v_i)|$. Every root chosen in this segment belongs to
    $\Anc(v_j)$ and appears no later than the first member of $\Anc(v_i)$. If
    $|\Anc(v_j)|\ge k$, Lemma~\ref{lem:ordering} places this member among the
    first $k$ vertices of $\Anc(v_j)$, so the segment meets at most $k$ parts;
    the same bound is immediate if $|\Anc(v_j)|<k$. Summing over the segments
    proves a bound of
    \[
      \lceil\log(2n^2)\rceil(1+\lfloor\log n\rfloor)
      =O(\log^2 n)
    \]
    parts for every path simultaneously.
  \end{proof}
\end{minipage}

\section*{Acknowledgements}
The author wishes to thank \'Etienne Bamas, Bundit Laekhanukit, and Shi Li for
fruitful discussions on the existence of subpolynomial approximations for related problems.

\section*{AI methodology statement}
\label{sec:ai-methodology}

This work was developed in extensive collaboration with GPT-6 Astra. GPT was
involved throughout the research and preparation of the manuscript.
During the final proofreading stage all updates were performed manually
by the author.

The author initially used the model to investigate Sherali--Adams integrality
gaps for the color version of Submodular Orienteering, in which the reward
counts the distinct colors visited by a path. In this investigation, the model
found a non-constructive subpolynomial upper bound on the integrality gap
after one round of the hierarchy. The author extracted key ideas from this
output and recombined them with other goals, again with the help of GPT,
ultimately leading to the result presented in this paper.

After the first complete proof, the author extensively edited it for readability, simplifying
arguments and structure and improving terminology, notation, language, and the presentation
of the work's context.

The author takes full responsibility for the correctness, originality, and
integrity of the paper, including all mathematical claims and references.

\bibliographystyle{alpha}
\bibliography{partition_orienteering_algorithm}

\end{document}